\documentclass[aps,twocolumn,groupedaddress]{revtex4}

\usepackage{graphicx}

\usepackage{amsmath}
\usepackage{amssymb}
\usepackage{amstext}
\usepackage{amsxtra}
\usepackage{version}
\begin{document}
\newcommand{\Tp}{T$_{\mathrm p}$}
\newcommand{\Tc}{T$_{\mathrm C}$}
\newcommand{\TN}{T$_{\mathrm N}$}
\newcommand{\Tco }{T$_{\mathrm CO}$}
\newcommand{\sinth}{\mbox{$\sin\theta/\lambda$}}
\newcommand{\inA}{\mbox{\AA$^{-1}$}}
\newcommand{\mub}{\mbox{$\mu_{B}$}}
\newcommand{\mns}{$-$}
\newcommand{\ddd}{3\textit{d}}
\newcommand{\pp}{2\textit{p}}
\newlength{\minusspace}
\settowidth{\minusspace}{$-$}
\newcommand{\msp}{\hspace*{\minusspace}}
\newlength{\zerospace}
\settowidth{\zerospace}{$0$}
\newcommand{\zsp}{\hspace*{\zerospace}}
\newcommand{\Cot}{$Co^{3+}$}
\newcommand{\Cof}{$Co^{4+}$}
\newcommand{\eg}{$\textit{e}_{g}$ }
\newcommand{\sqw}{S(\textit{Q},$\omega$)}
\newcommand{\Ts}{T$_{s}$\space}
\newcommand{\GG}{$\Gamma$\space}
\newcommand{\NaH}{Na$_{x}$CoO$_{2}\cdot y$H$_{2}$O}
\newcommand{\NaD}{Na$_{x}$CoO$_{2}\cdot y$D$_{2}$O}
\newcommand{\NaCo}{Na$_{x}$CoO$_{2}$}
\newcommand{\wat}{H$_{2}$O}
\newcommand{\deut}{D$_{2}$O}
\newcommand{\etal}{\textit{et al.}}
\newcommand{\degc}{$^{\circ}$C\space}
\newcommand{\br}{Br$_{2}$\space}
\newcommand{\aco}{$<Co-O>$}
\newcommand{\nax}{Na$_{x}$CoO$_{2}$}
\newcommand{\mco}{$\langle Co-O\rangle$}
\newcommand{\tm}{$T_{m1}$}
\newcommand{\tmm}{$T_{m2}$}


\title{Emergent charge ordering in near half doped Na$_{0.46}$CoO$_{2}$ }


\author{D. N. Argyriou}
\email[Email of corresponding author: ]{argyriou@hmi.de}
\affiliation{Hahn-Meitner-Institut, Glienicker Str. 100, Berlin D-14109, Germany}

\author{O. Prokhnenko}
\affiliation{Hahn-Meitner-Institut, Glienicker Str. 100, Berlin D-14109, Germany}

\author{K. Kiefer}
\affiliation{Hahn-Meitner-Institut, Glienicker Str. 100, Berlin D-14109, Germany}

\author{C. J. Milne}
\affiliation{Hahn-Meitner-Institut, Glienicker Str. 100, Berlin D-14109, Germany}

\date{\today}
\begin{abstract}
We have utilized neutron powder diffraction  to probe the crystal structure of layered Na$_{x}$CoO$_{2}$ near the half doping composition of $x=$0.46 over the temperature range of 2 to 600K. Our  measurements show evidence of a dynamic transition in the motion of Na-ions at 300K which coincides with the onset of a near zero thermal expansion in the in-plane lattice constants. The effect of the Na-ordering on the CoO$_{2}$ layer is reflected in the  octahedral distortion of the two crystallographically inequivalent Co-sites and is evident even at high temperatures. We find evidence of a weak  charge separation  into stripes of Co$^{+3.5+\epsilon}$ and Co$^{+3.5-\epsilon}$, $\epsilon\sim$0.06$e$  below \Tco=150K. We argue that changes in the Na(1)-O bond lengths observed at the magnetic transition at \tm=88K reflect changes in the electronic state of the CoO$_{2}$ layer. 
\end{abstract}

\maketitle

\section{Introduction}

The alkali cobaltates Na$_{x}$CoO$_{2}$ have been the subject of intense interest as they are a rare example of competing interactions on a triangular lattice that can be easily tuned by chemical means. Varying the amount of Na ($x$) produces a rich phase diagram which exhibits  spin dependent thermopower ($x=$0.75)\cite{Wang:2003rj}, metal-insulator transitions ($x=$0.5)\cite{Foo:2004cx,Huang:2004cx}, antiferromagnetism and 5K superconductivity at $x=$0.3 for a hydrated compound\cite{Takada} . More recently it has been realized both experimentally and theoretically that the role of the Na ions goes beyond providing a simple means to electronically dope the CoO$_{2}$ layer\cite{Roger:2007lr,Zhou:2007fk,Marianetti:2007ly}. Rather, the ordering of Na-ions leads to a potential that perturbs the CoO$_{2}$ layer to produce strong electronic correlations\cite{Roger:2007lr}. The role of these correlations is still under investigation but it demonstrates that these materials can exhibit frustration in two different ways, one by the triangular topology of the CoO$_{2}$ layer and the other by the Na induced potential. 

\begin{figure}[tb!]
\begin{center}
\includegraphics[scale=0.2]{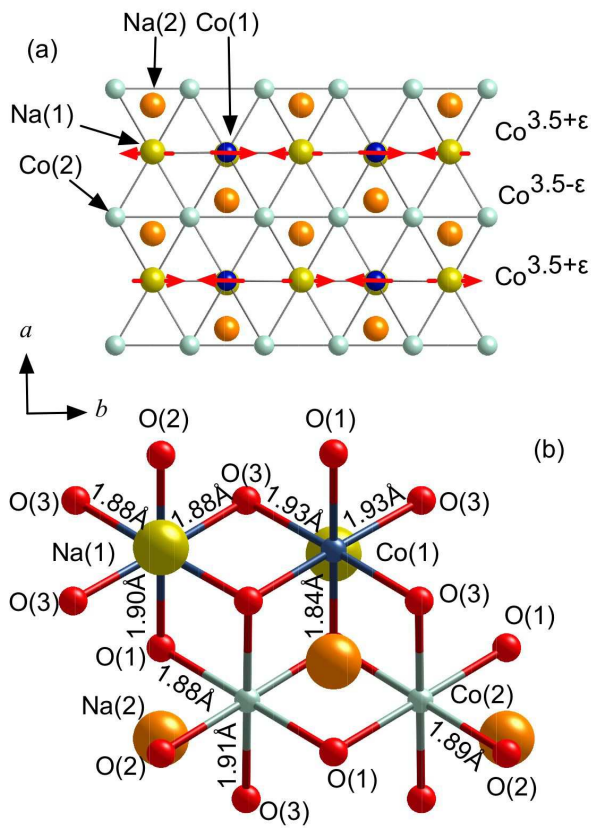} 
\caption{(Color online) Crystal structure of the Na$_{0.46}$CoO$_{2}$. (a) a projection of the $ab-$plane showing the crystallographically inequivelant Co(1) (dark blue) and Co(2) (light blue) sites and the  Na(1) (yellow) and Na(2) (orange) sites.  The Na-atom reside alternatively
above and below the CoO$_{2}$ sheet.  The Na(1) atom is directly above or below the Co(1) atom. The red arrows indicate the idealized  magnetic ordering of nominal low spin Co$^{4+}$ sites for the $x=$0.5 composition proposed on the basis of neutron and $\mu$SR measurements\cite{gasparovic}. (b) A portion of the CoO$_{2}$ sheet.   Note that the Na(2) site resides above a triangle defined by the edges of CoO$_{6}$ octahedra. Selected Co-O bond lengths determined at 2K are shown.}
\label{strt}
\end{center}
\end{figure}

This double frustration is best exhibited at half-doping. Here the Na ordering results in a relatively simple orthorhombic distortion of the parent hexagonal phase in sharp contrast to the complex incommensurate structures found for higher $x$ compounds\cite{Roger:2007lr}. For $x=$0.5 Na-ions order as to form stripes as shown in fig.\ref{strt} while the magnetic susceptibility shows two abrupt decreases (see for example inset in fig.\ref{powder}) at \tm=88K and at \tmm=52K\cite{Huang:2004cx,Foo:2004cx}.  The first transition is associated with the onset of a long range antiferromagnetic ordering\cite{Huang:2004cx,Foo:2004cx,gasparovic} while the second transition coincides with a sharp rise in the resistivity\cite{Huang:2004cx,Foo:2004cx}. This second transition has been ascribed to be driven by charge ordering (CO) of a $t_{2g}$ electron to  form distinct LS Co$^{3+}$ ($t_{2g}^{6}, S=0$) and LS Co$^{4+}$ ($t_{2g}^{5}, S=1/2$) ions \cite{Foo:2004cx}. Recent $\mu$SR and neutron diffraction measurements\cite{gasparovic,Mendels:2005cx,Yokoi:2005cx} propose a magnetic structure consistent with this picture, as the magnetic lattice  comprises of stripes of magnetically inactive Co$^{3+}$ and antiferromagnetic (AF) coupled Co$^{4+}$ (see fig.~\ref{strt}(a)). 

What is striking in this cobaltate is that the sequence of charge ordering and N\`eel transitions is reversed (\tmm$<$\tm), compared to classic charge ordered systems such as the manganites\cite{Argyriou:2000lx} or magnetite\cite{wright} and has brought the charge ordering  picture into some doubt.  To reconcile the fact that \tmm$<$\tm,  Bobroff \etal\cite{bobroff} propose on the basis of NMR measurements a scenario of a successive nesting of the Fermi surface (FS) that is coupled to a spin density wave in a way in which charge carriers are localized successively with decreasing temperature.  However, this idea has been more recently disputed as the role of the Na-ions and the crystal potential that they impose on the CoO$_{2}$ layer has been theoretically treated better. For example stripe Na ordering induces a weak incipient charge ordering on the CoO$_{2}$ layer\cite{Choy:2007ve,Zhou:2007fk,Marianetti:2007ly}, however the mechanism of a progressive nesting of the FS with decreasing temperature arises directly from the Na-ordering as shown by a Hubbard model using the spatially unrestricted Gutzwiller approximation\cite{Zhou:2007fk}. Here  the  charge ordering is viewed to be driven by the ordering of Na above the antiferromagnetic ordering at \tm=88K \cite{Choy:2007ve,Zhou:2007fk}.  Nevertheless this model successfully predicts the correct  antiferromagnetic ordering\cite{Choy:2007ve} and suggests a charge separation into stripes  of Co$^{+3.5+\epsilon}$ and Co$^{+3.5-\epsilon}$, with $\epsilon\sim$0.06$e$\cite{Zhou:2007fk}. This value represents  a very weak charge ordering and is below the lower limit of detection for charge separation by NMR\cite{bobroff}, but in agreement with powder diffraction measurements  that suggest  $\epsilon\sim$0.12$e$ at 10K\cite{williams}.

In this paper we use neutron powder diffraction (NPD) over a wide temperature range (2-600K) to probe the crystal structure of layered Na$_{x}$CoO$_{2}$ near the half doping composition of $x=$0.46 over the temperature range of 2 to 600K. Our NPD measurements show evidence of a dynamic transition in the motion of Na-ions at 300K which coincides with the onset of a near zero thermal expansion in the in-plane lattice constants of our Na$_{0.46}$CoO$_{2}$ sample. The effect of the Na-ordering on the CoO$_{2}$ layer is reflected in the  octahedral distortion of the two crystallographically inequivalent Co-sites and is evident even at high temperatures. We find evidence of a weak  charge separation  into stripes of Co$^{+3.5+\epsilon}$ and Co$^{+3.5-\epsilon}$, $\epsilon\sim$0.06$e$  below \Tco=150K, thus confirming a more physical sequence of charge ordering and magnetic transitions for this compound. We argue that changes in the Na(1)-O bond lengths observed at the magnetic transition at \tm=88K reflect changes in the electronic state of the CoO$_{2}$ layer. 

The paper is structured in the following way.  In section III we discuss the evidence for a weak charge ordering as determined from the temperature dependent NPD data.  The dynamic behavior of Na-ions at high temperature and changes in Na-O bond lengths close to the magnetic transition \tm\ are discussed in section IV, while the unusual temperature dependence of the lattice constants in section V. Discussion and summary are found in section VI and VII respectively. 

\section{Experimental}

\begin{figure}[bt!]
\begin{center}
\includegraphics[scale=0.15]{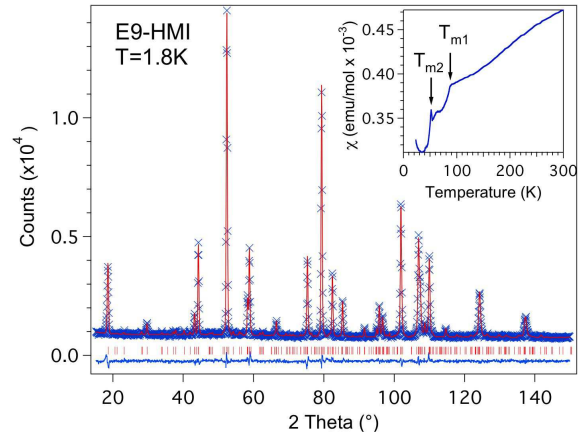} 
\caption{Rietveld refinement of NPD data measured at 1.8K from our $x$=0.46 sample.  Here crosses represent the measured NPD data, while the continuous line through the points represents the calculated diffraction pattern.  The difference between the data and the model is shown at the bottom of the figure.  Vertical bars represent expected Bragg reflections for the P$nmm$ structure of this compound.  The weighted R-factor here is $wR_{p}$=4.22\% and the Bragg R factor 4.3\%. In the inset we show the magnetic susceptibility of our Na$_{0.46}$CoO$_{2}$ powder sample measured on a SQUID magnetometer under a field of 1kOe. Arrows indicate two anomalies that have been ascribed to \tm=88K and \tmm=52K.}
\label{powder}
\end{center}
\end{figure}

Polycrystalline samples were prepared using standard solid state synthesis techniques.   The starting stoichiometry for these samples was Na$_{0.75}$CoO$_{2}$.  In order to deintercalate Na from the lattice to achieve a $x\sim$1/2 composition, a 5g portion of the $x=$0.75 sample was immersed in a bromine-acetonitrile solution with a 1:1 Na to Br$_{2}$ ratio, stirred in solution for 7-14 days and washed.  The Na/Co ratio of the product was measured using neutron activation analysis (NAA) giving a composition $x=$0.46(1). Magnetic susceptibility as a function of temperature ($\chi(T)$) was measured using a Quantum design MPMS and was found to be identical to the published literature as shown in the inset of fig. 2 \cite{Foo:2004cx,Huang:2004cx}. Rapid measurements of high resolution neutron powder diffraction data were collected from the $x$=0.46(1) sample using the HRPD diffractometer ($\Delta d/d\sim5\times 10^{-4}$) at the ISIS-facility, Rutherford Appleton Laboaratory. Higher statistics data suitable for Rietveld refinement were measured between 2 to 600K using the high resolution powder diffractometer E9 ($\Delta d/d\sim2 \times 10^{-3}$, $\lambda$=1.7973\AA), located at the Berlin Neutron Scattering Center, at the Hahn-Meitner-Institut (HMI).   Supplementary temperature dependent data were also measured from the $x=$0.75 sample between 5-300K. All NPD data were analyzed using the Rietveld method which allowed us to measure lattice parameters, atomic positions and atomic displacement parameters as a function of temperature. A typical Rietveld refinement of the NPD data is shown in fig.~\ref{powder}.

\section{Evidence of Weak Charge Ordering}

\begin{figure}[tb!]
\begin{center}
\includegraphics[scale=0.22]{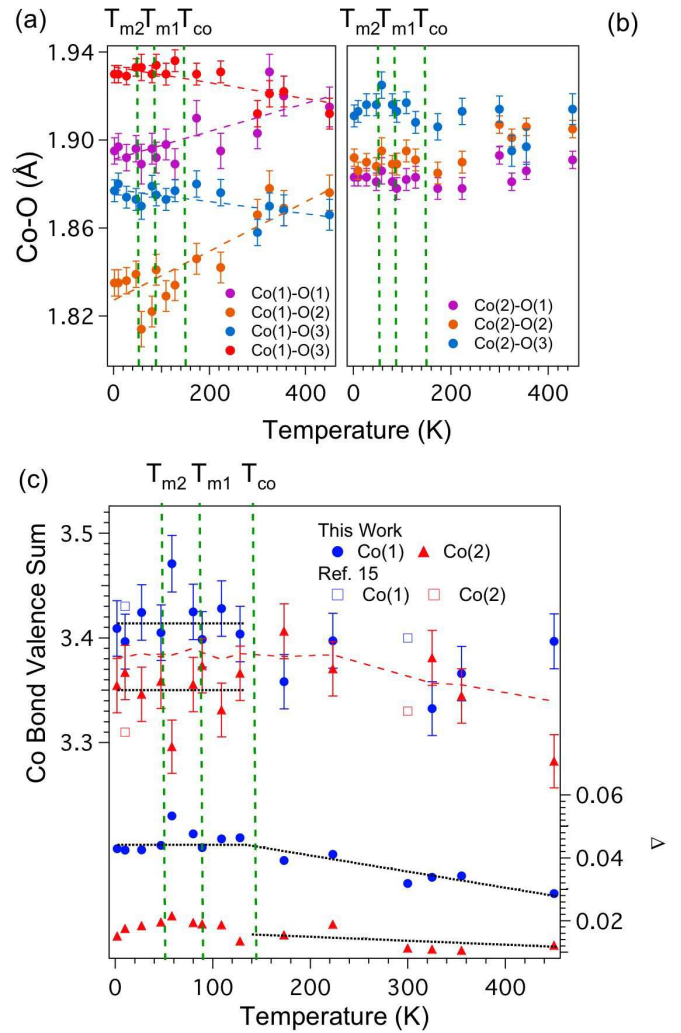} 
\caption{(Color online) (a,b) Temperature dependence of Co-O 
bond lengths computed from Rietveld analysis of NPD data measured on the E9 diffractometer.(c) Bond valence sums for the Co(1) and Co(2) sites (filled circles) as a function of temperature computed from Co-O bond lengths determined from Rietveld analysis of the NPD data.  The dashed line through the data is the average BVS for the two Co sites. BVS values obtained from ref. \onlinecite{williams} are also shown. On the same figure we plot the octahedral distortion parameter $\Delta$ for the CoO$_{6}$ octahedra centered in the Co(1) and Co(2) sites. Dashed lines are guides to the eye.}
\label{bonds}
\end{center}
\end{figure}

The validity of the reported orthorhombic P$nmm$ structure for $x=\frac{1}{2}$\cite{Huang:2004cx,williams} was tested by considering space groups that arise from distortions of the parent hexagonal structure P$6_{3}/mmc$ which are consistent with the reported orthorhombic unit cell. The program ISODISPLACE was used for this purpose\cite{cam}.  This approach for the symmetry analysis resulted in  space groups that were either centered or primitive monoclinic, in both cases  incompatible with the diffraction data.  Alternately, using as a starting point the reported space group P$nmm$, and testing for related space groups that are compatible with the diffraction  data resulted in primitive non-centrosymmetric solutions such as P$mm2$ or P$mm2_{1}$. Rietvelds analysis on the basis of these space groups resulted in somewhat poorer fits than the reported structure. Our best modeling of the NPD data  between 2-450K  were  obtained using the P$nmm$ model \cite{Huang:2004cx,williams}, producing refinements with $wR_{p}$ of approximately 4\% (between 2- 450K).  Contrary to the observation for higher $x$ compositions, we find no evidence of incommensurability in the NPD data\cite{Roger:2007lr}.  

The crystallography of the P$nmm$  structure has been published elsewhere\cite{Huang:2004cx,williams}, however for clarity we illustrate it in fig.~\ref{strt} and remind the reader that here there are two symmetry in-equivalent sites for Na (labeled Na(1) and Na(2)) and  two for Co (labeled Co(1) and Co(2)).  The local symmetry for the Co(1) and Co(2) sites differs in that the Na(1)-atom lies directly above or below the Co(1)-atom while the Na(2)-atom resides above a space formed between CoO$_{6}$ octahedra as illustrated in fig.~\ref{strt}(b). 

The effect of the Na-ion potential correlates with the distortion of the CoO$_{6}$ octahedra. This distortion is shown in fig.~\ref{strt}(b) where we illustrated the difference on the CoO$_{6}$-Na coordination and selected Co-O bond lengths.  Here the Na-ion potential results in a distortion of the Co(1)O$_{6}$ octahedron, so at 1.8K the six equal Co-O bonds found in metallic $x=0.75$ distort to form three long bonds (1.90-1.93 \AA) and three shorter bonds (1.84-1.88 \AA). In sharp contrast the Co(2)O$_{6}$ octahedron is more regular with bond lengths varying between 1.88 to 1.91 \AA. This indicates that the Co(1) ions experience essentially a different crystalline potential than the Co(2) ions, as a direct consequence of the Na ordering. In fig.~\ref{bonds}(a,b) we show the temperature dependence of the Co-O bond lengths for the two Co sites.  From these data it is evident that the larger distortion of the Co(1)O$_{6}$ octahedron is maintained from low until high temperature, while the spread of bond lengths for the Co(2)O$_{6}$ is smaller and relatively temperature invariant. These octahedral distortions can be quantified by the  parameter $\Delta$, where $\Delta=\sqrt\frac{\Sigma(\langle Co-O\rangle-(Co-O)_{i})^{2}}{\langle Co-O\rangle^{2}}$, $\langle Co-O\rangle$ is the average Co-O bond length and the summation is done over the 6 Co-O bonds. Here  $\Delta$ would be zero for a regular octahedron with 6 equivalent Co-O bonds.   We find that the Co(1)O$_{6}$ octahedron  is  more distorted by a factor of 3 at high temperatures (see fig.~\ref{bonds}(c)), compared to  the octahedron centered on the Co(2) site. With decreasing temperature however this difference increases to a factor of 5 while the  distortion saturates below $\sim$150K. As the data indicate on the same figure, the distortion of the Co(2) octahedron is much less sensitive to temperature.

At first sight, comparison of these bond lengths at 1.8K to the ideal LS Co$^{3+}$ and Co$^{4+}$-O bond lengths of 1.93 and 1.83\AA\space respectively would suggest that there is no evidence for integer charge separation between Co(1) and Co(2) sites. The analysis of the experimentally determined Co-O bond lengths using the bond valence sum (BVS) method allows us to estimate the difference in the valance of the two Co-ions. For the calculation of the BVS we used $b=$0.37 and $R_{o}=$1.70 \footnote{These values correspond to Co$^{3+}$-O bonds. Reliable values of Co$^{4+}$-O bonds are not available}\cite{Brese:st0462}. The BVS as a function of temperature for the two Co sites is shown in fig.~\ref{bonds}(c). For high temperatures we find that the BVS shows some scatter that  reflects changes in the mobility and ordering of Na atoms between CoO$_{2}$ sheets (see below) and possibly to geometrical differences that arises from the Na-ion potential imposed on the CoO$_{2}$ sheet over the same temperature range.\footnote{Overall the validity of the BVS method may not hold in the case of dynamic effects.  At high temperatures Na is mobile which is reflected in a large Debye-Waller factor  ($U_{iso}\sim30\times 10^{-3}\AA^{2}$) at around 300K which decrease smothly and rapidly to values similar as those found for Co and O ($U_{iso}\sim7\times 10^{-3}\AA^{2}$) below 100K.}  However, below $\sim$150K we find a small but measurable and consistent  difference of $\epsilon\sim$0.06$e$ in the BVS for the two Co atoms. Although the difference is comparably smaller than what is found in conventional charge ordered systems, the separation of the data into two values (one low and one high) below 150K is statistically significant. These data would suggest that below \Tco=150K  there is a separation of charge into Co$^{3.5+\epsilon}$ and Co$^{3.5-\epsilon}$ stripes running along the $b-$axis as shown in fig.~\ref{strt}(a). This charge ordered structure is in agreement with the magnetic neutron diffraction measurements, where the magnetically active Co-site would correspond to the Co(1) site with the slightly higher BVS and octahedral distortion. The values of the BVS obtained are in good agreement with both theoretical predictions\cite{Zhou:2007fk} and recently reported values at 10 and 300K respectively\cite{williams} which are also plotted on fig.~\ref{bonds}(c) for comparison. The weak charge ordering found here is consistent with the relatively low resistivity of this material at low temperatures ( $\sim 100\ m\Omega\ cm$ at 2K)\cite{Huang:2004cx,Foo:2004cx,gasparovic}. That the average BVS is approximately 3.3$e$ reflects the mixed valent nature of this compound.\footnote{The average BVS value lower than 3.5 reflect the absence of reliable parameters for Co$^{4+}$. A similar BVS number are noted in ref. \onlinecite{williams}}.

\section{Na Ordering and behavior of Na-O bond lengths}

\begin{figure}[tb!]
\begin{center}
\includegraphics[scale=0.3]{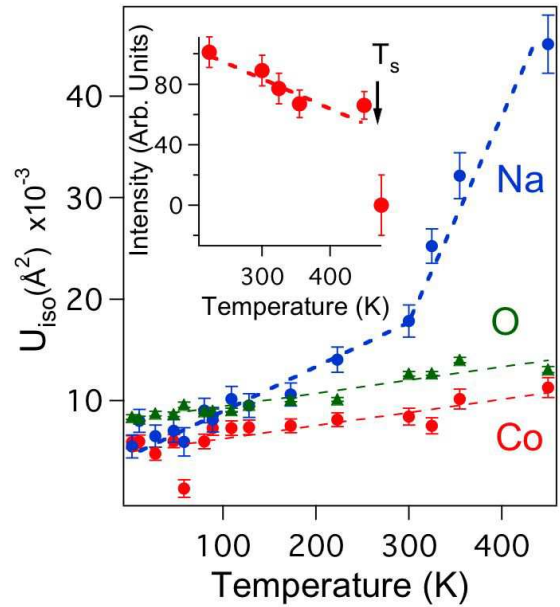} 
\caption{
(Color online) Isotropic atomic displacement parameters ($U_{iso}$) for the Na, Co and O-atoms determined from the Rietveld analysis of the NPD data.  In the analysis the following constraints were used $U_{iso}$(Co(1))=$U_{iso}$(Co(2)), $U_{iso}$(Na(1))=$U_{iso}$(Na(2)),$U_{iso}$(O(1))=$U_{iso}$(O(2))=$U_{iso}$(O(3)). Lines through the data are guides to the eye.  A slope change in the $U_{iso}$(Na) is evident around 300K. In the inset we show the temperature dependence of the (111) reflection in the orthorhombic P$nmm$ setting.  This reflection is a superlattice reflections with respect to the parent P$6_{3}/mmc$ crystal structure and arises from the ordering of Na ions.  The temperature $T_{s}\sim$460K marks the decomposition of the sample in hexagonal Na$_{x}$CoO$_{2}$ and Co$_{2}$O$_{3}$. }
\label{uiso}
\end{center}
\end{figure}

We now turn our attention to the behavior of the Na-layer for this composition.  In fig.~\ref{uiso} we plot the temperature dependence of the atomic displacement parameters $U_{iso}$ (Debye-Waller factor) determined from our Rietveld analysis.  We find that the $U_{iso}$ values of the O- and Co-atoms to be in general of the expected amplitude and show a linear behavior with temperature.  The behavior of $U_{iso}$ for the Na-ions however is unusual in that there is a clear change in slope at 300K separating a low  and a high temperature behavior, while for T$>$300K the $U_{iso}$ values for Na become  large. Such behavior is indicative of a dynamical transition occurring at 300K involving only the motion of Na-ions, as similar signatures are absent for the Co- and O-atoms\cite{Huang:2004prb}. Indeed such large vales of $U_{iso}$ suggest that Na-ions may become mobile between CoO$_{2}$ layers above 300K. 

For higher temperatures we find that our sample decomposes at $T_{s}\sim$460K to \nax\ and Co$_{2}$O$_{3}$. This transition is quantified by tracking the intensity of the (111) reflection as shown in the inset of fig.~\ref{uiso}. This reflection is a superlattice reflection with respect to the parent P$6_{3}/mmc$ structure and arises from the ordering of Na-ions \cite{Huang:2004cx}. Our neutron powder data measured at 475K indicate the loss of this and other superstructure reflections and a return to P$6_{3}/mmc$ symmetry with the addition of Co$_{2}$O$_{3}$ reflections. 

\begin{figure}[tb!]
\begin{center}
\includegraphics[scale=0.15]{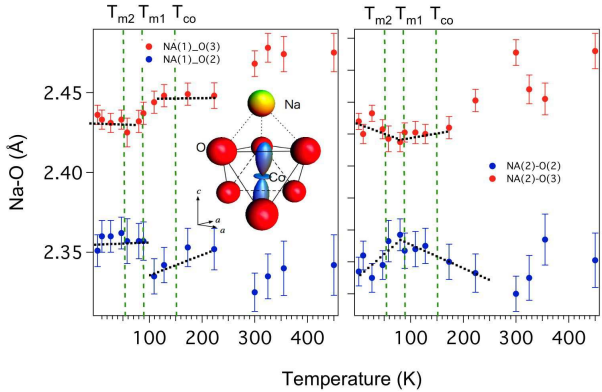} 
\caption{ Temperature dependence of and Na-O bond lengths computed from Rietveld analysis of NPD data measured on the E9 diffractometer. Dashed lines are guides to the eye.}
\label{Na_bonds}
\end{center}
\end{figure}

Within this perspective we now look more closely to the temperature dependence of the Na-O bond lengths show in fig.~\ref{Na_bonds}(a,b).  For both Na sites the Na-O bond lengths show a set of short bonds ($\sim$2.36\AA) and a set of long bonds ($\sim$2.44\AA). While the high temperature behavior is complicated by the high Na-ion motion as discussed above,  on cooling below 300K the long bonds decrease, while the short bonds show an increase down to 150K.  This correlated behavior in general indicates a displacement of Na-ions along the $a-$axis. At 150K we find that the small charge disproportionation  in the CoO$_{2}$ layer is not reflected in the Na-O bonds. Surprisingly however we find that at \tm\ a decrease of $\sim$0.01\AA\ of the long Na(1)-O bond and a correlated increase in the short Na(1)-O bond, while for the Na(2)-O bond lengths an opposite and less clear effect can be seen in the data.

We interpret these changes in the Na-O bond lengths as reflecting changes in the electronic state of the CoO$_{2}$ layer at \tm. The nature of any coupling between changes in the electronic state of the CoO$_{2}$ and Na can arises from the orbital configuration of the Co-ion itself. It is argued by Kroll \etal \cite{Kroll} that the edge-sharing Co$^{4+}$O$_{6}$ octahedra are compressed along the $c-$axis  reduces the point group symmetry to  $D_{3d}$. The $t_{2g}$ orbital of the Co$^{4+}$ is split in $D_{3d}$ as $t_{2g}$ = $a^{\prime}1_{g}$ + $e^{\prime}_{g}$, giving a fully occupied $e^{\prime}_{g}$ and a half filed $a^{\prime}1_{g}$. The latter orbital looks like a $3d_{z^{2}-r^{2}}$ orbital and points along the $c-$axis\cite{Kroll}. For the case of the higher valent Co(1)-ion (nominally Co$^{4+}$), this orbital would point  in between the O-atoms and towards the Na(1)-atom as shown in the inset of fig.~\ref{Na_bonds}(a). Since the charge disproportionation here is small each Co ions will have a similar electronic and orbital configuration. 

Therefore the orbital configuration of the Co-ions provides a means to couple electronically the CoO$_{2}$ and Na layers. The nature of the coupling is electrostatic and would arise from the occupation of the $e^{\prime}_{g}$ orbital. We would expect that changes in the electronic configuration of the CoO$_{2}$ sheet to be reflected also in the relative positions of the Na-ions as indicated by the Na-O bond lengths. More precisely changes in the electronic state of Co should be clearest for the Na(1)-O bonds as the Na(1)-ion sits directly above (or below) a Co(1)-ion. The same argument would suggest a less pronounced effect for the Na(2)-O bond lengths as the Na(2) ion resides above and between CoO$_{6}$ octahedra. This is indeed  reflected in the bond length data where a strong response is found in the Na(1)-O bonds and a less clear responce in the Na(2)-O bonds\footnote{The Na(2)-O response may arise from the repulsion of Na-ions}. At \tmm\ we find no clear evidence of changes in the Na-O bond lengths. This is expected as the changes in charge separation at this lower transition are computed to be much smaller than those at \tm\cite{Zhou:2007fk}.

\section{Anomalous behavior of Lattice Constants}

\begin{figure}[htb!]
\begin{center}
\includegraphics[scale=0.2]{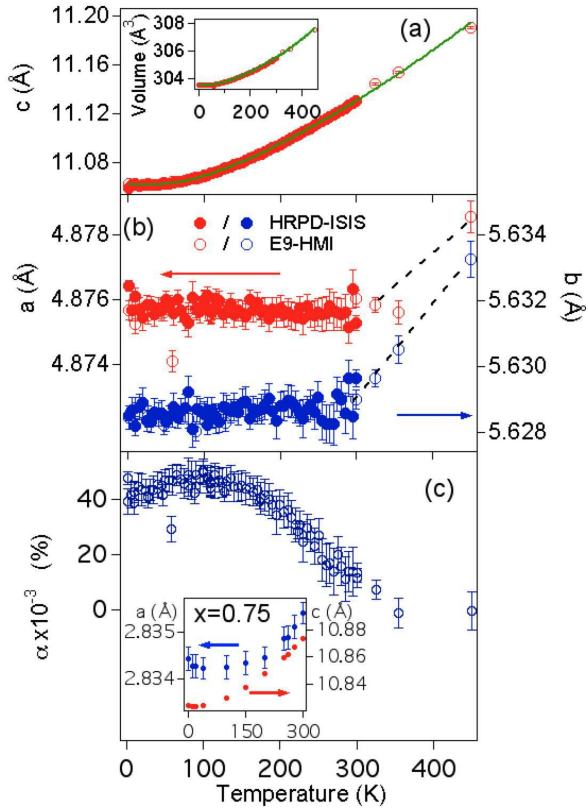} 
\caption{(Color online) (a,b) Lattice constants and unit cell volume (inset in panel (a)) obtained from the Rietveld refinement of NPD data measured on HRPD at ISIS (filled circles) and E9 (open circles) over the temperature range of 2 to 450K.  In panel (a) and inset the solid lines represent the thermal expansion obtained from a fit to the data using the second order Gr\"uneisen approximation.(c) Temperature dependence of the additional component to the thermal expansion $\alpha$ for Na$_{0.46}$CoO$_{2}$ obtained by subtracting the temperature behavior of the in-plane lattice constant of the $x=$0.75 compound. In the inset of panel (c) we show lattice parameters measured as a function of  temperature from a hexagonal $x=$0.75 sample. }
\label{lps}
\end{center}
\end{figure}

In  fig.~\ref{lps}(a-b) we show the temperature dependence of the lattice constants determined from Rietveld refinement of the NPD data. These data show a positive thermal expansion (TE) for the $c-$axis between 2 and 450K, but for the $a-$ and $b-$axis we find an almost constant TE between 2 and 300K; here for $T<$300K linear TE expansion coefficients are $-9(3)\times^{-7}$/K and $1.1(4)\times^{-6}$/K for $a$ and $b$ respectively. Such small TE was also discussed in ref. \onlinecite{williams} for a much more limited number of temperatures and smaller range in temperature. For $T>300$K there is a return to positive TE for both in-plane parameters. This crossover  coincides with the dynamic transition in the motion of the Na-ions as indicated by the behavior of $U_{iso}$ for Na.
 
The behavior of the lattice constants for this $x=$0.46 sample is in sharp contrast to our $x=$0.75 sample (shown in the inset fig.~\ref{lps}), were we find a positive TE for both $a-$ and $c-$axes between 5 and 300K. Assuming that TE is dominated by acoustic phonons below 300K for Na$_{x}$CoO$_{2}$ materials  and whose frequency is relatively invariant between $x=$0.75 to $x=$0.46, the $T-$dependence of the lattice constants for the $x=$0.75 sample (see inset in fig.~\ref{lps}(f)) can be used to quantify the difference in the in-plane TE between these two samples.  Here we define the term  $\alpha=(a'-a_{hex})/a'$ where $a'=(\sqrt{3}a+b/2)/2$, $a$ and $b$ are the orthorhombic  lattice constants of the $x=$0.46 compound and $a_{hex}$ is the hexagonal lattice constant of the $x=0.75$ compound.\footnote{Here $a'$ was normalized to equal $a_{hex}$ at 300K.}  For the higher temperature  data (T$>$300K) $a_{hex}$ was assumed to vary linearly with temperature. Here $\alpha$ represents an additional temperature dependent contribution to the expected in-plane lattice constants (as defined by $a_{hex}$) and its temperature dependence in shown is fig~\ref{lps}(c).  For T$>$ 300K $\alpha$  is near zero, although the data in this region are more limited.  However for T$<$300K  $\alpha$ increases with decreasing temperature and reaches a value of $\sim40\times10^{-3}\%$ at 1.8K.  The maximum value of $\alpha$ is $\sim50\times10^{-3}\%$ at 150K, close to \Tco.  The decrease below this temperature is due to an increase in the $a_{hex}$ for the $x=$0.75 as seen in the inset of fig.~\ref{lps}(c). We speculate that the anomalous TE below 300K may be driven by electronic correlations induced by Na-ions.

\section{Discussion}

The measurement we present here suggest a picture of incipient charge ordering for near half doped Na$_{x}$CoO$_{2}$. At high temperatures the ordering of Na-ions defines two different CoO$_{6}$ octahedra, one that is relatively undistorted and one that more distorted.   On cooling the differences in octahedral distortions between these two different Co-sites becomes larger and may reflect an increasing influence of the Na-ion potential on the CoO$_{2}$ sheet, as Na ions become more localized around their mean crystallographic positions.
Indeed  the near zero TE for the in-plane lattice constants coincides with a dynamical transition in the displacement parameter $U_{iso}$ of the Na-ion. Further the  physical meaning of $\alpha$ can be  interpreted as a measure of the Na-induced electronic correlations onto the CoO$_{2}$  layer which in this view saturate at \Tco=150K.

While our NPD work can correlate the distortion of the CoO$_{6}$ octahedra and the Na-ordering even at high temperatures, it is not until \Tco\ that we find evidence for a weak charge separation into stripes. Indeed it is possible that this pattern of charge ordering is present in the lattice from the onset as a direct result of the Na-ordering but it in effect is hidden by the dynamic behavior of the Na-ions. Therefore we argue that the charge ordering emerges at low temperatures as Na-motion becomes  more confined. 

More critically in terms of the physics of these materials we demonstrate that charge ordering occurs at a higher temperature that the magnetic ordering and electronic transitions at \tm\ and \tmm\ respectively. This is consistent with recent theoretical models that suggest that the Na potential imposes a degree of charge ordering to the lattice\cite{Choy:2007ve,Zhou:2007fk,Marianetti:2007ly}. Indeed Zhou and Wang  \cite{Zhou:2007fk}, suggest that as much as half of the expected  charge disproportionation  would occur at a temperature above \tm\footnote{We compute these change by averaging over site 1 and site 3 in reference \onlinecite{Zhou:2007fk}}, consistent with our observations.  

At lower temperature the changes in the CoO$_{2}$ layer may be inferred indirectly by monitoring  the Na(1)-O bonds. Here the orbital configuration of the Co provides for charge density pointing directly to the Na(1)-ions thus providing a sensitive parameter to electronic changes in the CoO$_{2}$ layer.  Indeed changes in the Na-O bond lengths may be more sensitive than changes in Co-O bonds as $e_{g}$ axial orbitals are empty. Our measurements find that that changes in the Na(1)-O bond lengths correlate with the magnetic transition at  \tm\ suggestive of further changes in the electronic state of the CoO$_{2}$ layer.  It is predicted that charge separation is enhanced gradually below \tm\cite{Zhou:2007fk} but the changes here are overall again small and may fall outside the limits of our sensitivity. At \tmm\ we find no evidence of changes in the lattice or changes in the lattice symmetry.  The prediction of a modulation of the amplitude of antiferromagnetically coupled spins  as well as the charge within a Co$^{+3.5+\epsilon}$ stripe is much smaller than our detection limit ($\sim$0.02$e$)\cite{Zhou:2007fk}. 

The structural observations at \Tco\ we report here correlate with features in the charge dynamics.  For example Quian \etal\space\cite{qian:046407}  report from ARPES measurements that with increasing $T$ from the insulating region (were a clear gap is found) the size of the gap and the spectral weight around the gap decrease. Although the gap closes at \tm\space the spectral weight does not completely vanish until approximately 120K, a behavior that is  attributed to the formation of quasiparticles that gain significant weight due to coupling along the $c-$axis. For similar temperatures optical spectroscopy measurements find a broad feature that is associated with fluctuating charge ordering or a CDW in both anhydrous\cite{Wang:2004cx,jhwang} and hydrated superconducting samples\cite{lemmens:167204}. These observation together with our structural measurements point  towards a picture  where at 150K an incipient charge ordering forms. 
 
\section{Summary}


In summary this work establishes that $(a)$ the  Na-ordering on the CoO$_{2}$ layer is reflected in the  octahedral distortion of the two crystallographically in-equivalent Co-sites and is evident even at high temperatures; $(b)$ The charge ordering occurs below \Tco=150K, a temperature higher than the magnetic ordering found at  \tm=88K, consistent with theoretical models that suggest that the Na potential imposes a degree of charge ordering to the lattice\cite{Choy:2007ve,Zhou:2007fk,Marianetti:2007ly}; $(c)$ Below \Tco\ we find a weak charge ordering into stripes of Co$^{3.5+\epsilon}$ and Co$^{3.5-\epsilon}$ with a $\epsilon\sim$0.06 $e$, a value in good agreement with that obtained from a Hubbard model using the Gutzwiller approximation\cite{Zhou:2007fk}; $(d)$ A dynamic transition in the motion of Na-ions occurs at 300K and coincides with the onset of a near zero thermal expansion for the in-plane lattice constants of our Na$_{0.46}$CoO$_{2}$ sample.

\begin{acknowledgments}
The authors thank P.G. Radaelli,  and L.C. Chapon for helpful discussions and W.S. Howells for assistance in the collection and reduction of the HRPD data. 
\end{acknowledgments}


\end{document}